\documentclass[fleqn,10pt]{wlscirep}
\usepackage{aas_macros}
\usepackage[utf8]{inputenc}
\usepackage[english]{babel}
\usepackage{amsmath}
\usepackage{amsfonts}
\usepackage{amssymb}
\usepackage{graphicx}
\usepackage{lmodern}
\usepackage[T1]{fontenc}
\usepackage[scaled=0.9]{DejaVuSans}
\usepackage{hyperref}

\usepackage{csquotes}
\usepackage[style=nature, 
			backend=biber%
		   ]{biblatex}
\newcommand{\ntheta}{n^0}
\newcommand{\nphi}{n^1}
\newcommand{\Nside}{$N_\text{side}$}

\title{Morphometry on the sphere: Cartesian and irreducible Minkowski tensors explained and implemented}

\author[1,*]{Caroline Collischon}
\author[2,3,4]{Michael A. Klatt}
\author[5]{Anthony J. Banday}
\author[1]{Manami Sasaki}
\author[2,4]{Christoph Räth}
\affil[1]{Dr. Karl-Remeis Sternwarte and Erlangen Centre for Astroparticle Physics, Friedrich-Alexander Universität Erlangen-Nürnberg, Sternwartstr. 7, 96049 Bamberg, Germany }
\affil[2]{Deutsches Zentrum f\"ur Luft‐ und Raumfahrt (DLR), Institut f\"ur KI Sicherheit, Wilhelm‐Runge‐Stra\ss{}e 10, 89081 Ulm, Germany}
\affil[3]{Deutsches Zentrum f\"ur Luft- und Raumfahrt (DLR), Institut f\"ur Materialphysik im Weltraum, 51170 K\"oln, Germany}
\affil[4]{Department of Physics, Ludwig-Maximilians-Universit\"at M\"unchen, Schellingstr.~4, 80799 Munich, Germany}
\affil[5]{IRAP, Université de Toulouse, CNRS, CNES, UPS, Toulouse, France}

\affil[*]{caroline.collischon@fau.de}

\begin{abstract}
  Minkowski tensors are comprehensive shape descriptors that robustly
  capture $n$-point information in complex random geometries and that
  have already been extensively applied in the Euclidean plane. Here, we
  devise a novel framework for Minkowski tensors on the sphere. We first
  advance the theory by introducing irreducible Minkowski tensors, which
  avoid the redundancies of previous representations. We, moreover,
  generalize Minkowski sky maps to the sphere, i.e., a concept of local
  anisotropy, which easily adjusts to masked data.
  We demonstrate the power of our new procedure by applying it to
  simulations and real data of the Cosmic Microwave Background,
  finding an anomalous region close to the well-known Cold Spot.
  The accompanying open-source software, \texttt{litchi}, used to
  generate these maps from data in the HEALPix-format is made publicly
  available to facilitate broader integration of Minkowski maps in
  other fields, such as fluid demixing, porous structures, or
  geosciences more generally.
\end{abstract}

\begin{document}

\flushbottom

\maketitle

\thispagestyle{empty}

\section*{Introduction}

Minkowski functionals (MF) and Minkowski tensors (MT) from integral
geometry~\cite{chiu2013,schneiderweil2008} are powerful and versatile
shape descriptors for random spatial structures in real space. They
provide a localized and comprehensive shape analysis by characterizing,
among others, symmetries and preferred directions. They provide robust
access to information from $n$-point correlation functions (since they
can be expressed as a sum over all $n$-point correlation
functions)~\cite{kerscher01}. Yet, their calculation is much simpler
than that of higher-order correlation functions, which allows for
an exploitation of higher-order correlations in spatial structures that is
practically inaccessible when relying on estimates of correlation
functions.  

MT have been used in a broad range of fields, such as analysis of
trabecular bone structure
\parencite{rath2008}, classifying the shapes of galaxies
\parencite{rahman2004}, fluid demixing \parencite{bobel2016}, analyzing
cellular, granular, and porous structures \cite{schroederturk2011},
source detection in gamma-ray astronomy \cite{goring2013,Klatt2019},
analyzing nuclear pasta matter \cite{schuetrumpf2015} or applications in
crystallography (\parencite{bobel2018} and references therein). However,
most applications have so far focused on the Euclidean space.

A prominent exception is cosmology, specifically the analysis of simulated \cite{Rocha_2005,Elsner_2009} and reconstructed maps \cite{Planck_IX_2016} of the
Cosmic Microwave Background (CMB). The scalar MF have already been
intensively used to analyze the CMB, either using single MF
\cite{Gott1989,Colley1996} or in systematic studies of all MF
\cite{schmalzing1998,ducout2012,DuqueMarinucci2024}. They have been used to search for
non-Gaussianity in temperature and polarization, both by the Planck
collaboration \parencite{planck2016-l07} and other groups
\cite{komatsu2010,modest2013,carones2023,chingangbam2023}.
Rank 2 MT have also been used by \cite{joby2019,chingangbam2017},
finding consistency with Gaussianity via their approach.

In cosmology, MF and MT offer several advantages over other methods,
such as two- or three-point correlation functions. 
Their definition in real space offers a more
comprehensive, ``human-readable" way to analyze all-sky data as compared
to harmonic decomposition. Since they can be calculated locally for
either single shapes or selected regions, MF/MT are excellent tools to
search for anisotropy and provide natural ways to deal with
incomplete (e.g., masked) data. Additionally, they contain higher-order
information that would not be included in a 2- or 3-point correlation
analysis.

Here, we introduce a pixel-based approach of calculating MF/MT of
arbitrary rank and implement it in the first publicly available code
for a Minkowski analysis on the sphere\footnote{
\url{https://github.com/ccollischon/litchi}}. Unlike the
aforementioned references where the MF/MT were calculated globally, we devise
a localized analysis, dubbed Minkowski Maps (MM).

We start by defining the classical Cartesian MF/MT on the sphere and
show how to use scalars to represent their information about the degree
of anisotropy for rank 2 and 4. Next, we define, for the first time,
irreducible MT on the sphere, providing access to scalar shape
information at arbitrary rank. The MM visualize local information. We
provide methods to calculate these maps on the sphere and apply them to
examples (which can serve as a reference for users). Note that in
contrast to MM in Euclidean space a MM on the sphere requires the
concept of parallel transport on the sphere, as explained below.
Finally, we apply our methods to Planck CMB temperature data and
simulations. Importantly, our methods provide additional information
compared to the more common tools as they do not strongly react to
absolute values of a random field but rather to its shape and
symmetries. Thus, we find two noteworthy spots, one in the southern
Galactic hemisphere close (but not statistically related) to the Cold
Spot \cite{vielva2004}, and the other above the Galactic plane.

\section*{Theory of MT: Euclidean plane vs sphere}

In the Euclidean plane, the three MF are given by area, perimeter, and
Euler characteristic. In general the MF of a compact body $K$ with a
smooth contour $\partial K$ are defined as

\begin{equation}
W_0(K) = \int_K \text d r\, , ~ W_\nu(K) = \int_{\partial K} G_\nu \text d r
\end{equation}
where $\nu \in \{1,2\}$, $G_1 = 1$ and $G_2 = \kappa$, the sectional curvature.
The definition can be straightforwardly extended to more general
domains, e.g., polyconvex sets~\cite{schroederturk2010,schneiderweil2008}.

The MF are additive ($W_\nu(A) + W_\nu(B) = W_\nu(A\cup B) - W_\nu(A\cap
B)$). In fact, in Euclidean space, Hadwiger's theorem states that any
additive functional of convex shapes that is continuous and motion
invariant (i.e., scalar) can be expressed as a linear combination of MF
\parencite{hadwiger1957}. In that sense, MF capture all `additive shape
information.'

On the sphere, an analogous theorem holds as
proven by Klain and Rota \cite[Theorem 11.3.1]{klainrota1997}.
In particular, shapes in pixel images can be expressed as unions of
single (convex) pixels, making them a good target for MF/MT analysis.
Grayscale images with varying brightness can be analyzed by applying a
threshold, interpreting everything above the threshold as part of the body, and
optionally adding up the MF/MT obtained at several thresholds.

The analysis with MF can be naturally generalized for anisotropic
structures to motion covariant tensors, the so-called MT, which are defined in Euclidean
space as follows. Using the symmetric tensor product with 

\begin{equation}
\mathbf{n}^b := \underbrace{\mathbf{n}\otimes \ldots\otimes \mathbf{n}}_{b\text{ times}}
\end{equation}
and components 

\begin{equation}
(\mathbf{n}^b)_{i_1\ldots i_{b}} =n_{i_1} \ldots\, n_{i_{b}}\, .
\end{equation}
The translation invariant Cartesian MT of rank $b$ are then given by

\begin{equation}
W_\nu^{0,b}(K) := \int_{\partial K} \mathbf{n}^b G_\nu\, \text d r
\end{equation}
where different communities may use different (constant) prefactors.
There are also translation covariant MT, which we omit here since
there is no generic generalization to the sphere.
Note that $W^{0,1}_\nu(K) = 0$ for closed $K$ in flat space. Alesker
proved a theorem analogous to Hadwiger for MT \cite{alesker1999}.
However, such a theorem is unknown on the sphere. Further properties of
the MT (useful for applications) can, e.g.,
be found in \cite{schroederturk2010,schroederturk2011,schroederturk2013}.

\begin{figure}
\centering
\includegraphics[scale=1.3]{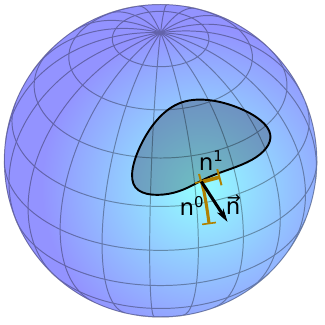}
\caption{Definition of $\mathbf{n}$ on the sphere. It is perpendicular
to the contour of the body to be analyzed (black) and is decomposed into
its components on the spherical surface (orange). No embedding into the
three-dimensional space takes place.}
\label{fig:sphere}
\end{figure}

On the sphere, $\mathbf{n}$ is interpreted as the tangent space vector
perpendicular to the contour. It is normalized according to the metric
of the spherical surface, $g_{00}=1\, ,~ g_{11}=\sin^2(\theta)\, ,~
g_{01} = g_{10} = 0$. We will subsequently use the notation $n^\mu$ for
its components with $\mu\in \{ 0,1\}$, $\ntheta=n^\theta;\,
\nphi=n^\phi$ in the usual coordinates on the sphere; see
Fig.~\ref{fig:sphere}. Tensor integration on the sphere can be done
according to \cite{fabian1957}, who presents an expansion method
analogous to Taylor expansions.

\subsection*{Scalar representation based on Cartesian representation}

To visualize and analyze the degree of anisotropy with MT, they need to
be brought into a scalar form that is invariant under rotation. 
Several possible representations are available in the Euclidean plane
that can be generalized to the sphere. Each option captures different
information. The trace operator is a straightforward choice. For
$W_1^{0,2}$, we obtain  

\begin{equation}
\mathrm{tr}\left(W_1^{0,2}(K)\right) = \int_{\partial K} (\mathbf{n}\otimes \mathbf{n})^{\mu\nu}g_{\nu\mu} \text d r = \int_{\partial K} (\ntheta)^2 + \sin^2(\theta)\,( \nphi)^2 \text d r = \int_{\partial K}\text d r = W_1^{0,0}(K)~ ,
\end{equation} 
where we inserted the metric tensor. 

More morphological information is contained in the ratio of eigenvalues,
i.e., the eigenvalue quotient (EVQ). 
The eigenvector equation for eigenvalue $\lambda$ for a vector $v^\mu$ is given by

\begin{align}
 W^{\mu\nu}g_{\nu\alpha}v^\alpha &= \lambda v^\mu \\
\begin{pmatrix}
W^{00} & W^{01}\sin^2(\theta)\\
W^{10} & W^{11}\sin^2(\theta)\\
\end{pmatrix}
\begin{pmatrix}
v^0 \\ v^1
\end{pmatrix}  &= 
\lambda
\begin{pmatrix}
v^0 \\ v^1
\end{pmatrix}
\end{align}
where $W^{\mu\nu} = \left( W_1^{0,2} \right)^{\mu\nu}$ are the components of $W_1^{0,2}$. 

Since $W_1^{0,2}$ is symmetric, the eigenvectors are orthogonal. For an
elongated shape with 2-fold symmetry, one eigenvalue will be larger than
the other so that the ratio is a useful measure for elongation. In this
paper, we divide the larger eigenvalue by the smaller one. The
eigenvectors themselves can be used to find the preferred orientation of
the body.

An approach to rank 1 tensors was shown by \cite{joby2021}, 
who integrated over the gradient of the field. In our framework, this
tensors corresponds to $W_1^{0,1}$, where the contour is given by contours of the
field at several thresholds. This tensor vanishes for closed contours in
Euclidean space. When looking at a non-closed shape in a small enough
region to keep spherical effects negligible, its length measures the
directionality of the contours (how much the remaining gradient prefers
one direction). 
Alternatively one can use the direction of that vector to show the total
normal of the remaining contours.


Rank 4 tensors can be tackled by taking their symmetries into account. We will
do this along the lines of a method shown by \cite{mehrabadi1990}, where the authors
considered three-dimensional rank 4 tensors as six-dimensional rank 2 tensors
(that is, 6x6 matrices) using their symmetries. In this representation,
eigenvalues can be easily calculated, and the corresponding eigenvectors
can then be reinterpreted as eigentensors in the original space.

In our case, a symmetric two-dimensional rank 4 tensor can be represented by a
3$\times$3 matrix whose eigenvectors are reinterpreted as rank 2 tensors.
Multiplication of the rank 4 tensor with a rank 2 tensor $\sigma^{\mu\nu}$ is given in components by

\begin{align}
\sigma'^{\mu\nu} &= W^{\mu\nu\alpha\beta} g_{\alpha\gamma} g_{\beta\delta}\, \sigma^{\gamma\delta} \\
\sigma'^{00} &= W^{0000}g_{00}g_{00}\sigma^{00}+ 2 W^{0001}g_{00}g_{11}\sigma^{01} + W^{0011}g_{11}g_{11}\sigma^{11} \\
\sigma'^{01} = \sigma'^{10} &= W^{0100}g_{00}g_{00}\sigma^{00}+ 2 W^{0101}g_{00}g_{11}\sigma^{01} + W^{0111}g_{11}g_{11}\sigma^{11} \\
\sigma'^{11} &= W^{1100}g_{00}g_{00}\sigma^{00}+ 2 W^{1101}g_{00}g_{11}\sigma^{01} + W^{1111}g_{11}g_{11}\sigma^{11}
\end{align}
where the factor 2 comes from the symmetry of all involved tensors. Note
that $W$ has only 5 distinct components (no change under permutation of indices). Inserting the components of $g$
and writing in matrix form gives the equation for eigentensor $\sigma$ with eigenvalue $\lambda$ as

\begin{equation}
\label{eq:rank4mat}
\begin{pmatrix}
 W^{0000} & 2 W^{0001}\sin^2(\theta) & W^{0011}\sin^4(\theta) \\ 
 W^{0100} & 2 W^{0101}\sin^2(\theta) & W^{0111}\sin^4(\theta) \\ 
 W^{1100} & 2 W^{1101}\sin^2(\theta) & W^{1111}\sin^4(\theta)
\end{pmatrix}
\begin{pmatrix}
\sigma^{00} \\ 
\sigma^{01} \\ 
\sigma^{11}
\end{pmatrix}
=
\lambda
\begin{pmatrix}
\sigma^{00} \\ 
\sigma^{01} \\ 
\sigma^{11}
\end{pmatrix}
\end{equation}
The eigenvalues of this 3$\times$3 matrix can be used for further analysis.

\subsection*{Irreducible Minkowski tensors}

The redundancy of the Cartesian representation can be entirely avoided by the
so-called irreducible Minkowski tensors (IMT). This representation
expands the MT in spherical harmonics so that a tensor of rank $s$ only
captures $s$-fold anisotropy. In other words, the MT are decomposed
according to their symmetry.

In Euclidean space, they are obtained by calculating a normal vector
density for a shape. This density describes what fraction of the body's
normal vectors point in which direction. In the plane, these directions are
distributed on $[0,2\pi)$. The Fourier coefficients of the density of
normal vectors are called IMT; see~\cite{klatt_mean-intercept_2017,
collischon2021, klatt_characterization_2022} for two dimensions (2D),
and \cite{Kapfer2011, kapfer_jammed_2012, mickel_shortcomings_2013} for
three dimensions (3D). The IMT are complex scalars. They provide a
natural way to characterize both the degree of anisotropy and
directional information for any rank via their phase and absolute value.
Such a scalar representation is essential for higher orders, where
Cartesian tensors become unwieldy. 

Here, we define IMT for the first time on the sphere. Note that a simple
embedding in 3D fails to distinguish isotropic from anisotropic bodies
on the sphere. For example, a spherical cap is an isotropic body on the
sphere, but IMT of an embedding in 3D would, undesirably, assign to it a high degree
of anisotropy. The solution is to parallel transport the
normal vectors to a single point, where they can then be treated as in the
Euclidean plane (i.e., calculating their angular density). This 
suitable representation provides a powerful characterization of spherical
anisotropy, as demonstrated below. Hence, this definition is an important
step towards an effective shape analysis on the sphere for anisotropies
and symmetries of arbitrary rank.

Let $\rho_K$ be the normal density of a body $K$, where the
functional value $\rho_K(\varphi)$ is proportional to the portion of the
body's contour that points into the direction $\varphi$. 
The total integral equals the contour length of the boundary $\partial K$. 
For a spherical polygon (such as any shape defined by pixelated
images on a sphere), the normal density can be written as

\begin{equation}
\rho_K(\varphi) = \sum_k L_k\,\delta(\varphi-\varphi_k)
\end{equation}
where the edges are indexed by $k$, have lengths $L_k$, and orientations
$\varphi_k$. On the sphere, $\varphi$ is determined from the direction
of the normal vector $\mathbf{n}$ parallel transported to a common
point. We choose $0$ to represent south, and $\pi/2$ east.

The IMT are then obtained by Fourier transforming $\rho_K$:

\begin{equation}
\psi_b(K) = \int_0^{2\pi} e^{ib\varphi}\rho_K(\varphi)\, \text d \varphi \stackrel{\text{polygon}}{=} \sum_k L_k e^{ib\varphi_k}
\end{equation}

The absolute values $|\psi_b(K)|$ of the IMT are then useful measures
for the degree of anisotropy with respect to $b$-fold symmetries.
Furthermore, the corresponding preferred directions are given by
$\varphi_n=(2\pi n + \arg(\psi_b))/b$ where $n \in \{ 0,\ldots,\,b-1
\}~$.

\section*{Calculating Minkowski maps}

\begin{figure}
    \centering
    \includegraphics[scale=0.7]{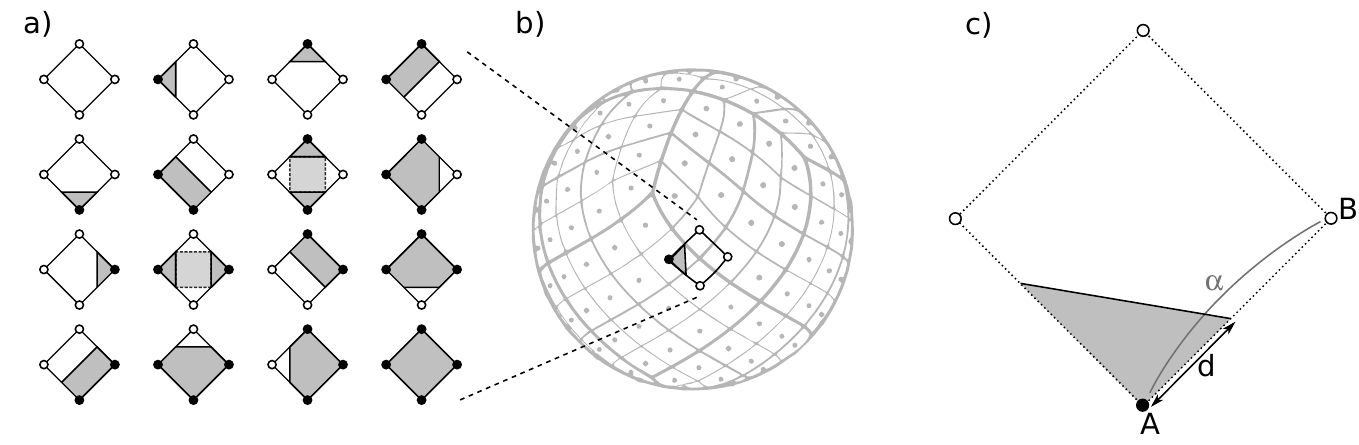}
    \caption{The computation of a marching square on HEALPix. a) Possible configurations of a 2$\times$2\,pixel-window.
    The small circles represent pixel centers that are either above
    (open) or below (filled) a certain threshold. The union of gray
    areas approximates the unpixelated shape within the observation
    window. In the diagonal cases, the corners are assumed to be
    disjoint if the total average lies above the threshold, and
    connected otherwise \parencite{appleby2018}. b) Position of a
    marching square window within the HEALPix grid (HEALPix background
    image vectorized from \cite{Martinez-Castellanos_2022}, Fig.~1c,
    \href{https://creativecommons.org/licenses/by/4.0/}{CC BY 4.0}). c)
    Interpolation of a contour segment (solid line) between pixel
    centers, where $A$ marks the center of a pixel below the threshold,
    and the other circles mark pixel centers above the threshold. The point
    where the interpolated contour lies between two pixel centers is
    calculated according to eq.~\ref{eq:interpolation}. Here, $\alpha$
    is the distance from $A$ to $B$ and $d$ the distance from $A$ to the
    interpolated contour in that direction.}
    \label{fig:marchingSquare}
\end{figure}

Localized MT for grayscale images are calculated in the
form of Minkowski maps (MM) using local observation windows. In
Euclidean space, code is available in the papaya2
\parencite{Schaller2020} and the related banana
\parencite{collischon2021} libraries, the latter refining the former for
astronomical image analysis. We adapt those techniques for squares in
the Euclidean plane on the sphere using the HEALPix
scheme\footnote{\url{https://healpix.sourceforge.io}}
\parencite{healpix}, see Fig.~\ref{fig:marchingSquare}. This scheme is
commonly used for CMB data and further described in the methods section.
The body to be described by the tensors is obtained by choosing a
brightness threshold, treating everything below as part of the body. Our
implementation, \texttt{litchi}, is available at
\url{https://github.com/ccollischon/litchi}.

Our scheme is based on a marching square algorithm. First, a window of
2$\times$2 pixels is selected by choosing all pixels that meet in one
corner, using the pixel centers as the corners of the window. (At
certain positions in HEALPix, only three pixels meet in a triangular
shape, which is taken into account accordingly.) All 2$\times$2
neighborhoods together form a marching square grid, as explained in more
detail in the methods section. Each pixel value is then either above or
below the given threshold, yielding 16 possible cases of black-and-white
neighborhoods, shown in Fig.~\ref{fig:marchingSquare} (left).

These configurations define how the contour segment passes through the
window, where we assume these segments to be geodesics. Their endpoints
at the edge of the window are interpolated between the neighboring pixel
centers depending on the threshold and pixel values using a technique by
\cite{mantz2008}, also shown in Fig.~\ref{fig:marchingSquare} (right).
The diagonal cases are chosen to be connected or disconnected based on
the total average \parencite{appleby2018}.

Let $\alpha$ be the distance between both pixel centers, denoted by $A$
and $B$. Let $t$ be the chosen brightness threshold,
and $p_{A}$ and $p_{B}$ be the pixel values at $A$ and $B$ respectively.
Then the distance $d$ from $A$ to the contour in the direction of $B$ is
given by 

\begin{equation}
\label{eq:interpolation}
d = \alpha\,  \frac{t - p_A}{p_B - p_A}.
\end{equation}

The normal vector pointing away from the body in tangential space is
then calculated for the window. Since these segments are short, a
zeroth-order approximation for the integral is used, where the length of
the segment is simply multiplied by the tensorial integrand
\parencite{fabian1957}. 

Note that the EVQ of $W_1^{0,2}$ of such a short straight line will
diverge as one of the eigenvalues goes to zero. To avoid this divergence
and, more importantly, to access structural information over an extended
area, we used larger windows on top of the 2$\times$2 marching squares. The
locally calculated segment-tensors are parallel transported to the
center of this enlarged window and then summed up. Each large window should
contain enough contour segments at various angles to allow for a
sensible analysis. Finally, one can use an output map at lower resolution to
depict the results, where pixel centers correspond to the centers of the
larger windows.

This procedure uses a single threshold, which may be an insufficient
representation of the structure. For a more comprehensive analysis,
we can add up tensors that are calculated at several thresholds before
smoothing to portray the data as accurately as necessary. Compared to a
previous method implemented by \cite{goyal2021}, where localized
structural information was only calculated at specific
pixelation-dependent scales and positions, MM provide a more versatile
approach. Further implementation details can be found in the
methods section.

\section*{Test cases}

For demonstration purposes, we generated MM for several exemplary shapes.
All shapes occupy only a few degrees around the equator to keep
spherical effects small and the results intuitive.

\subsection*{Rank 2}
\begin{figure}
\centering
\begin{tabular}{p{0.47\textwidth}p{0.49\textwidth}}
{\sffamily a) \hspace{0.15\textwidth} Cartesian}    & {\sffamily b)  \hspace{0.15\textwidth} Irreducible} \\
\includegraphics[width=0.47\textwidth]{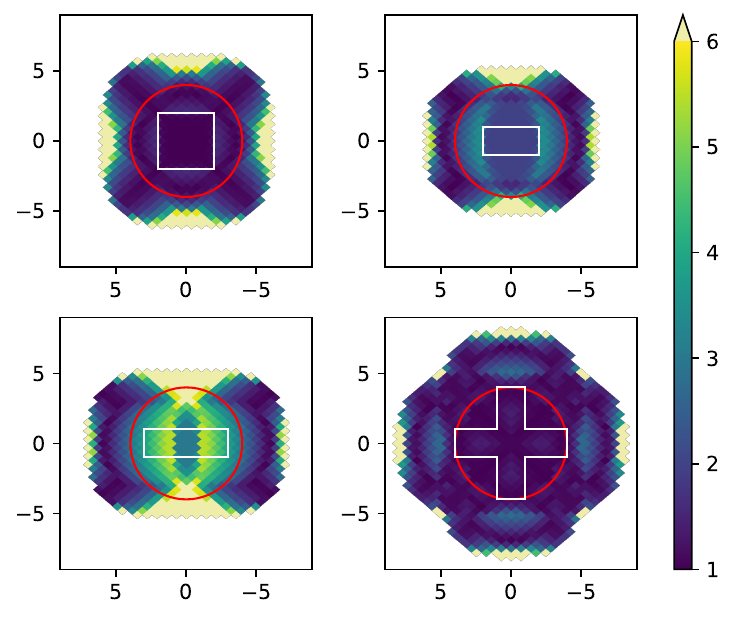} &
\includegraphics[width=0.49\textwidth]{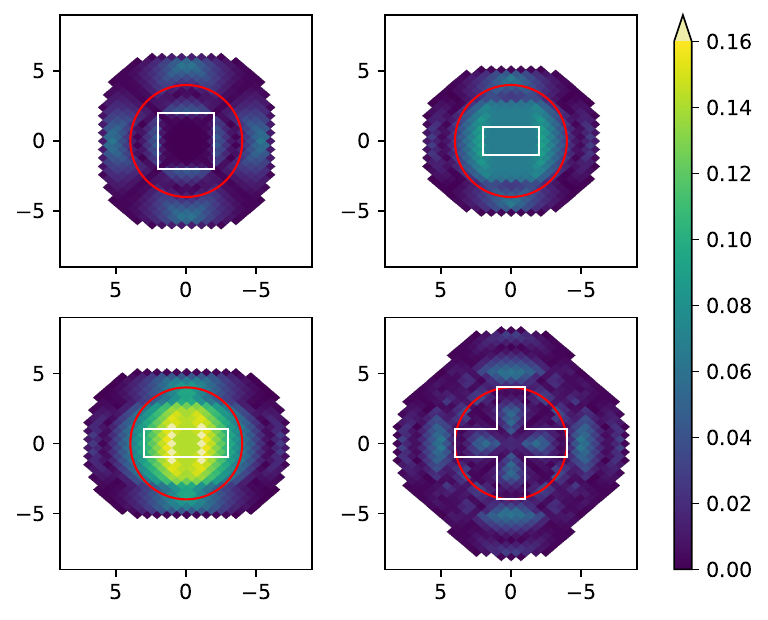}
\end{tabular}
\caption{Minkowski maps of rank 2 anisotropy measures of a
    square, elongated rectangle, longer rectangle, and cross. a) Using the EVQ of $W^{0,2}_1$.  b) Using $|\psi_2| $. Each white contour depicts the shape of the (input) body, and the red circles indicate the window size. Coordinates
    are given in degrees. (Input \Nside\ was 512 and that of the output 128.)
}
\label{fig:rectangles}
\end{figure}

We begin with the EVQ of $W_1^{0,2}$ and $|\psi_2|$ of a square and two
rectangles with aspect ratios 2:1 and 3:1 respectively. The
corresponding MM are shown together with the input shape and the moving
window size in Fig.~\ref{fig:rectangles} a) and b); see the panels at
the top, left and right, and the bottom left panels. Near the origin,
all shapes are entirely contained within the window, and the anisotropy
depends on the aspect ratio; the EVQ is an explicit function of the
aspect ratio. When the window is moved sideways, the respective opposite side
of the rectangle moves out of the window; hence, the anisotropy
increases. Even further away from the center, the anisotropy decreases
as the remaining sides at the top and bottom get shorter and cancel out
with the nearest edge. Finally, only one side (and increasingly shorter
sections thereof) is left in the window, leading to a diverging EVQ
while $|\psi_2|$ remains low. If only a corner is left in the window,
the EVQ is close to one, and $|\psi_2| $ is close to zero as two
perpendicular contours have no two-fold anisotropy. The same test was
done for a cross a) and b); see Fig.~\ref{fig:rectangles}, bottom right
panels. Now, the anisotropy is low as long as most of the shape is within
the window, except if only one bar is within the window. From that point
on, the map exhibits the same patterns as for the rectangles.

As expected, very high values in the Cartesian maps signify that only
a part of the shape is within the window. Larger windows relative to the
shapes smooth out these edge effects because the relative area
on the MM where the window contains the whole shape increases with the
window size, whereas regions where edge effects prevail depend on the
size of the shape. 

For an extended pattern such as the CMB temperature distribution, the
window size needs to be large enough compared to a characteristic length
scale to avoid diverging border effects. Even larger windows can be used
to check the relative alignment of the shapes making up the pattern.
Such a choice corresponds to analyzing the whole cross as opposed to a single
bar.

Care needs to be taken near masked regions where only a few contour
segments may be left in an otherwise sufficiently large window. We,
therefore, only allow a small fraction ($<1/16$) of input pixels in a
window to be masked, treating the whole window as masked otherwise. More
details on our masking can be found in the methods section.

\subsection*{Rank 4}

\begin{figure}
    \centering
    \begin{tabular}{p{0.47\textwidth}p{0.49\textwidth}}
    {\sffamily a) \hspace{0.15\textwidth} Cartesian}    & {\sffamily b)  \hspace{0.15\textwidth} Irreducible} \\
    \includegraphics[width=0.48\textwidth]{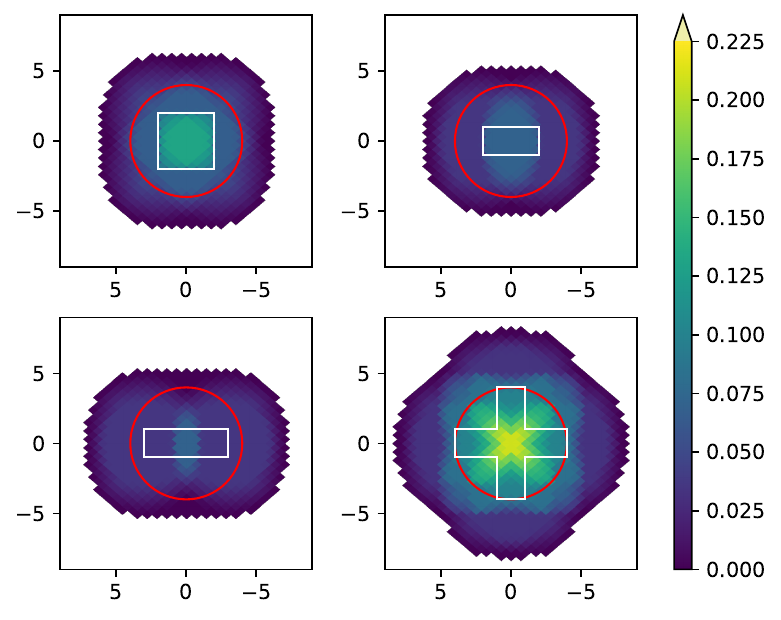} &
    \includegraphics[width=0.47\textwidth]{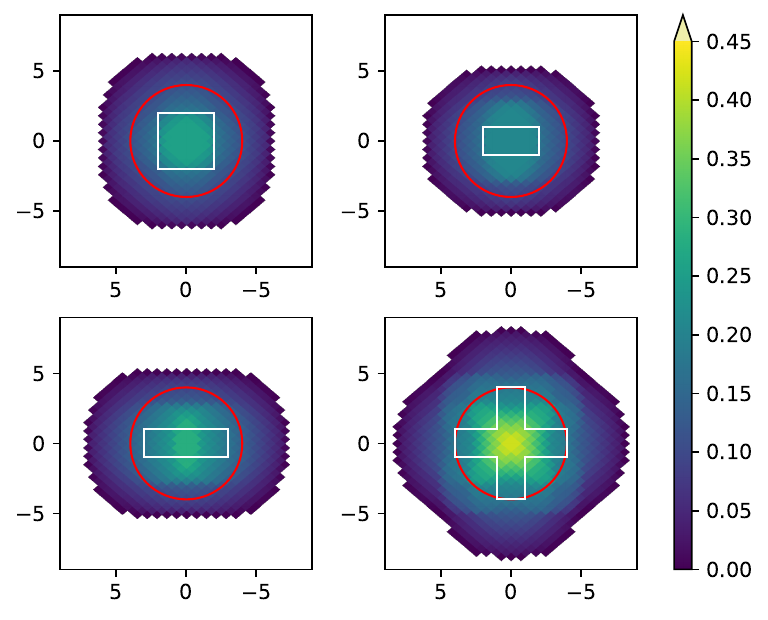}
    \end{tabular}
    \caption{Minkowski maps of rank 4 anisotropy measures of a
    square, elongated rectangle, longer rectangle, and cross. a) Using the mid eigenvalue of $W^{0,4}_1$. b) Using
    $|\psi_4|$. Each white contour depicts the shape
    of the (input) body, and the red circles indicate the window size.
    Coordinates are given in degrees.
     (Input \Nside\ was 512 and that of the output 128.)}
     \label{fig:rectangles_rank4_mid}
\end{figure}

We have tested various options to depict $W_1^{0,4}$ using the
eigenvalues of the matrix in eq.~\ref{eq:rank4mat}; for more details,
see the methods section. The most useful choice for our shape analysis
turned out to be the tensor's mid eigenvalue (i.e., its second
largest eigenvalue). Maps using this measure and $|\psi_4|$ of the same
example shapes as above are shown in
Fig.~\ref{fig:rectangles_rank4_mid}.

\subsection*{Threshold dependent graphs}
Additional information about the structure in a grayscale image with
more than just black-and-white pixels can be gained by varying the
brightness threshold, which changes the shape of the contours. The
chosen shape measure can then be plotted as a function of the brightness
threshold. This procedure is an established way to apply the MF/MT to
data that has been used, e.g., by \cite{planck2016-l07} or
\cite{joby2019}, as well as in
\cite{schroederturk2011, goring2013, klatt_characterization_2022}.











\section*{MT analysis applied to Planck data}

\begin{figure}
    \centering
    \includegraphics[width=\textwidth]{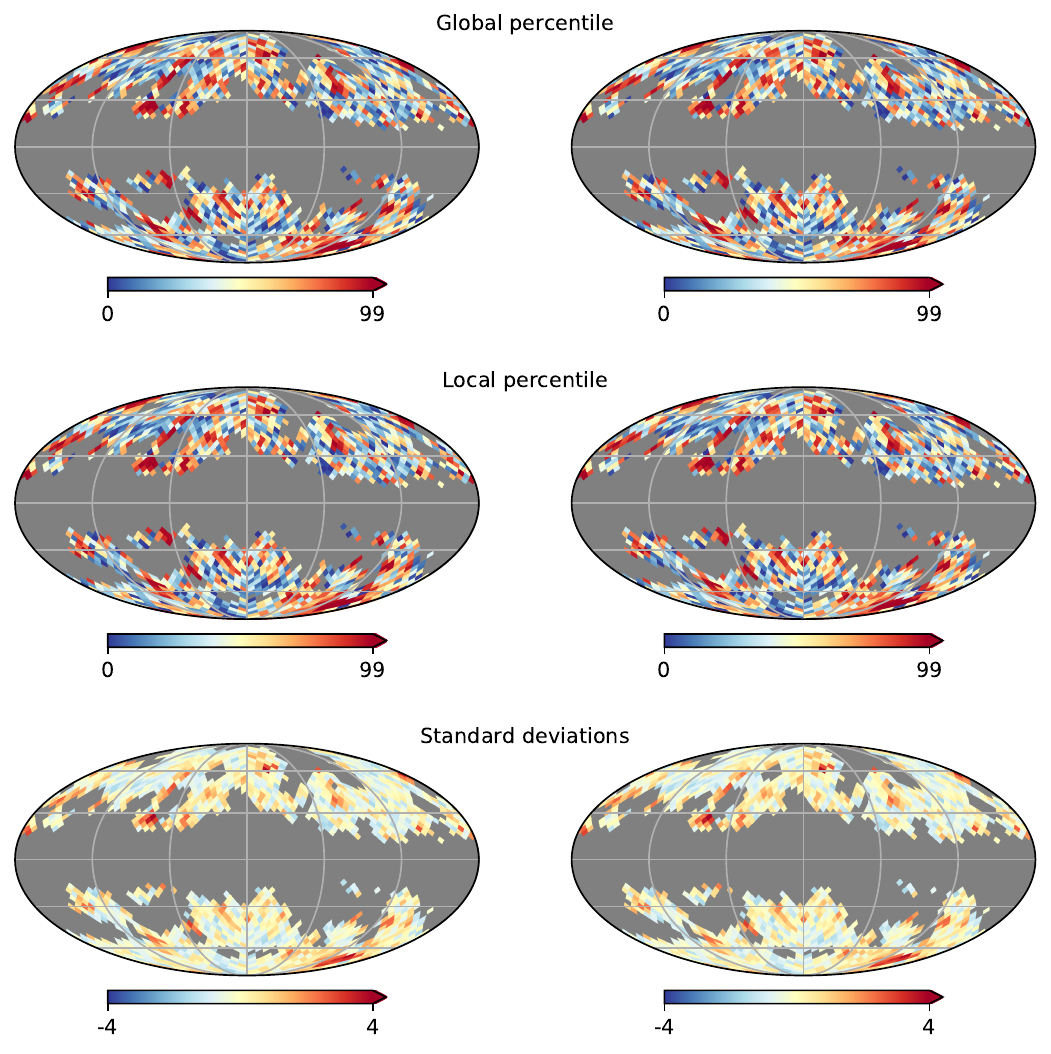}
    \caption{Deviation between SMICA data and simulations via the EVQ of $W^{0,2}_1$,
    window size 6$^\circ$, masked with the Common Mask, using various methods. Left: no smoothing of
    the MM, Right: MM smoothed with FWHM $6^\circ$. Top: global
    percentiles, Center: local percentiles, Bottom: multiples of local
    standard deviation }
    \label{fig:compSimuDataVariations}
\end{figure}

Next, we use our methods developed above to analyze real CMB data,
specifically Planck temperature data. More details and explanations
regarding the basics of Planck data products and analysis are given in
the methods section.

In our analysis, we calculate MM of the EVQ of $W^{0,2}_1$ for 13
thresholds between -3 and 3 standard deviations. All thresholds are
combined in each MM, in contrast to \cite{joby2019}. We use Planck PR3
CMB SMICA temperature data and 999 FFP10 simulations at \Nside= 512,
smoothed with FWHM$=20'$, and a MM window radius of $6^\circ$. We,
moreover, use the Common Mask (also at \Nside= 512 and smoothed with
FWHM$=20'$, setting all pixels above 0.95 to unmasked), and \Nside\ of
the output MM is 16. Since the MM window radius is larger than the
output pixel size, each output pixel describes its local surroundings,
avoiding divergences due to a lack of contours in the window. As above,
we took the (pointwise) mean and standard deviation of the simulation
MM and used these to estimate the local deviation from the data; see
Fig.~\ref{fig:compSimuDataVariations} (bottom). We also tested smoothing
the MM again, after calculating the scalar map, to check for
larger-scale deviations in anisotropy at this scale; see
Fig.~\ref{fig:compSimuDataVariations}, left for unsmoothed and right for
smoothed data. 

\begin{figure}
    \centering
    \includegraphics[width=0.8\textwidth]{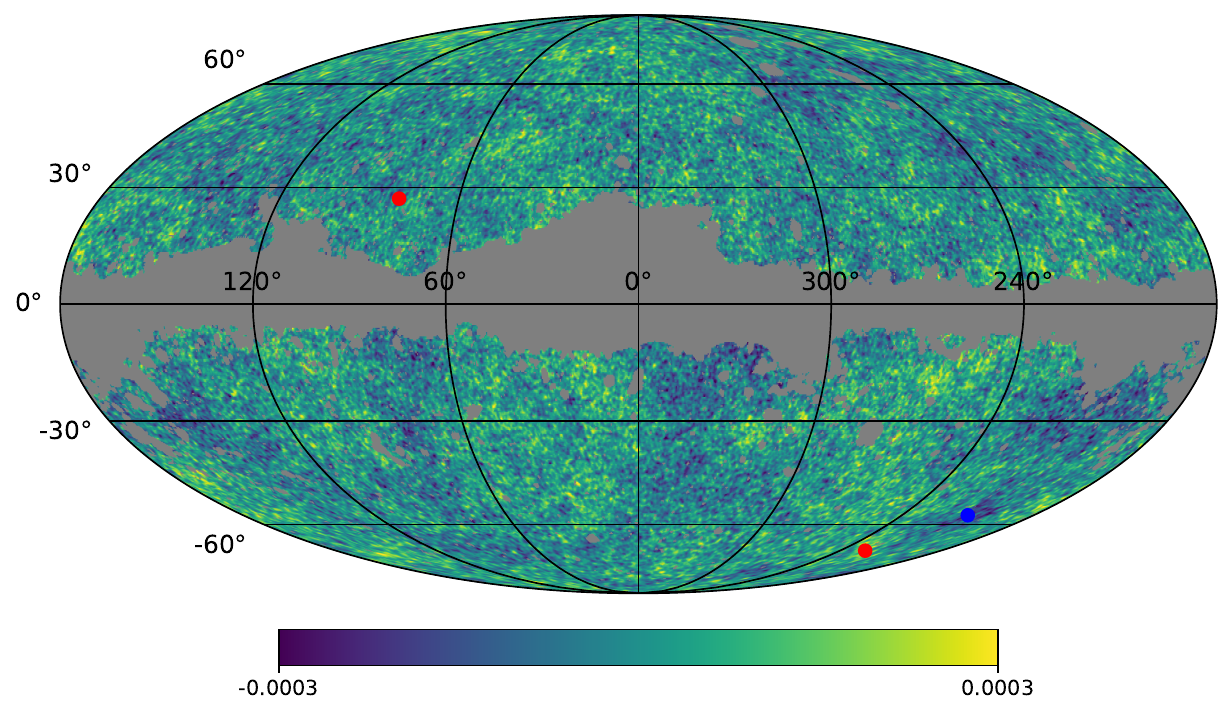}
    \caption{SMICA PR3 CMB temperature map and the detected anomalous
    regions (red dots) and the Cold Spot (blue dot).}
    \label{fig:smicaregions}
\end{figure}

Our analysis identifies two spots of interest: one close to
$(l_1,b_1)=(225,-69)$ and one near $(l_2,b_2)=(80,27)$, which are
highlighted by red marks in the temperature map of
Fig.~\ref{fig:smicaregions}. The first spot is close to the so-called
Cold Spot at around $(l,b)=(210,-57)$, an anomaly first detected in data
by the WMAP mission \cite{vielva2004,cruz2005,vielva2010} and later also
in Planck data \cite{planck2013-p09,planck2014-a18}. However, the
distance of $(l_1,b_1)$ to the Cold Spot is slightly larger than the
window diameter, which suggests that these spots are statistically
independent. For both $(l_1,b_1)$ and $(l_2,b_2)$, the data map shows a
deviation in anisotropy compared to the simulations by more than
$4\,\sigma$.

The MM pixel values do not follow a Gaussian distribution, so units of
standard deviation are not an ideal measure of deviation. Better
measures are the local and global percentile of the data and the
simulations, that is, the fraction of simulation MM pixels that take a
lower value than the given data MM pixel. These percentiles can be
evaluated either pointwise (comparing pixels at the same positions) or
globally (comparing each data pixel to all simulation pixels). The
results are shown in Fig.~\ref{fig:compSimuDataVariations} (center and
top rows). The local method has the advantage that it takes pixel distortion
into account, whereas the global method provides better statistics.
Using only the local method has the risk of comparing an outstanding
point in the data to an average point in the simulations, ignoring
similar points.
Both the local and global analyses find the same two spots at 
percentile values $>99\%$. 

The spots are also identified using PR3 data and simulations by the
Commander pipeline and when using the absolute values of irreducible
tensors of rank 2 instead of the EVQ of $W^{0,2}_1$. For Commander, the
standard deviation of the simulation MM pixels is larger, which naturally
results in a smaller multiple of standard deviation. However, we still
obtain high percentile values ($>97\%$ global smoothed MM, $>99\%$
global unsmoothed and local MM). Varying the window size by about 10\%
gives qualitatively the same results. The spot $(l_2,b_2)$ is located
near a masked region in the MM, which is due to several smaller masked
spots [see Supplementary Fig.~1 in SI and compare to Supplementary Fig.~2
depicting $(l_1,b_1)$], but we assume that those masked areas do not
significantly change the result since they are cut out rigorously.
Furthermore, the window contains enough contours to exclude edge
effects; see the methods section for further details on masking.

To see whether the analysis is affected by the aforementioned
non-contour-preserving properties of the HEALPix grid, both regions are
rotated to the center of the coordinate system, and the MM of the data
is recalculated. The deviation there is slightly smaller but still
remarkable ($>3.5\sigma$/$>99.4\%$ for $(l_1,b_1)$;
$>4.4\sigma$/$>99.9\%$ for $(l_2,b_2)$; using unsmoothed MM and for both
percentiles). 

Looking into those regions with threshold-dependent graphs using 600
simulations, we find that the southern spot at $(l_1,b_1) = (225,-69)$ exceeds 99\%
of the simulations (local percentile), whereas the other spot does not.
The excess in the map of all thresholds combined probably occurs due to
correlations between maps calculated for several different single
thresholds so that it exceeds in total but not for a single threshold. 

Note that, in a similar analysis, \cite{joby2019} also analyzed Planck
temperature data by calculating the eigenvalue ratio of $W^{0,2}_1$ with
a different method based on covariant derivatives of the field. They
took a global approach, integrating over the whole sphere at once at
various (single) thresholds ranging from -3 to 3 standard deviations of
brightness for each map. Thus, the EVQ can be plotted as a function of
the brightness threshold. Using the Planck CMB temperature map given by
the SMICA component separation pipeline and 100 FFP8 simulations at a
resolution of \Nside= 512 (all smoothed with FWHM=$20'$ and masked with
the Common Mask), they find good agreement between data and simulations. 

For a better comparison to \cite{joby2019}, we created threshold
dependent graphs of the whole sky. For this purpose, we used MM with a
window size of 6 and 10 degrees and averaged globally for each
threshold. The data is around the 1st percentile of the simulations,
making it more isotropic than the average simulation. These results are
clearly an interesting confirmation of isotropy and comparable to the
results of \cite{joby2019}, who found a similar alignment of structures
looking at only the 30\,GHz channel, probably due to beam effects.

To look into larger structures, we created MM with a window radius of
30 degrees and the same parameters as above. Due to masking, only caps
around the polar regions remain to be analyzed. The data stays below
98.5\%\ of the simulations both locally and globally.

\begin{figure}
    \centering
    \begin{tabular}{lll}
    {\sffamily a)} & {\sffamily b)} & {\sffamily c)} \\
    \hspace{-0.2cm}\includegraphics[trim=0.3cm 0 -0.3cm 0, clip, width=0.32\textwidth]{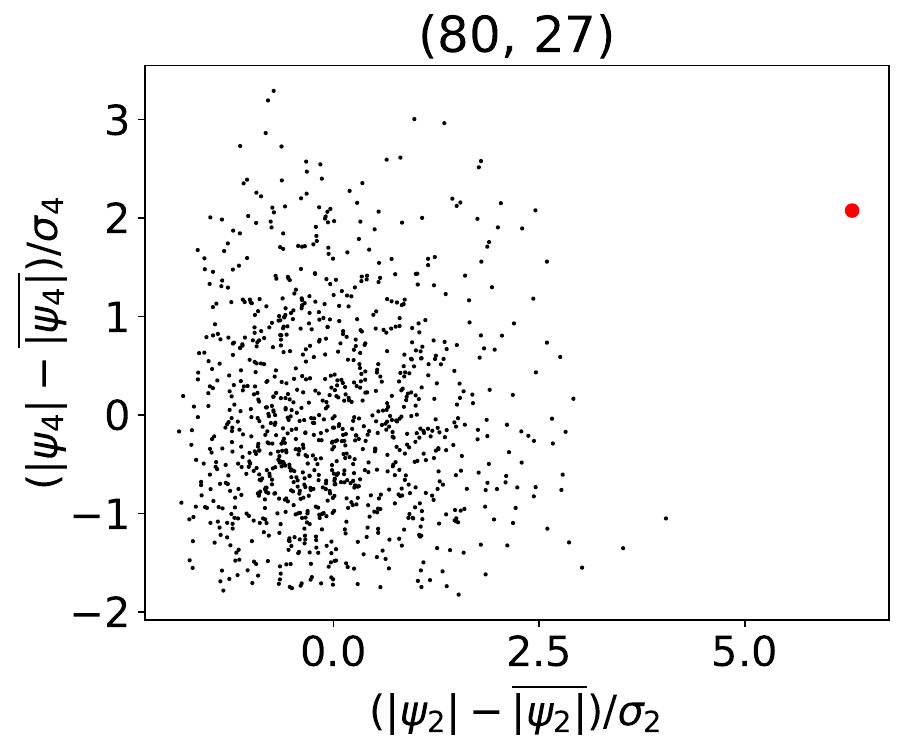}&
    \hspace{-0.2cm}\includegraphics[width=0.32\textwidth]{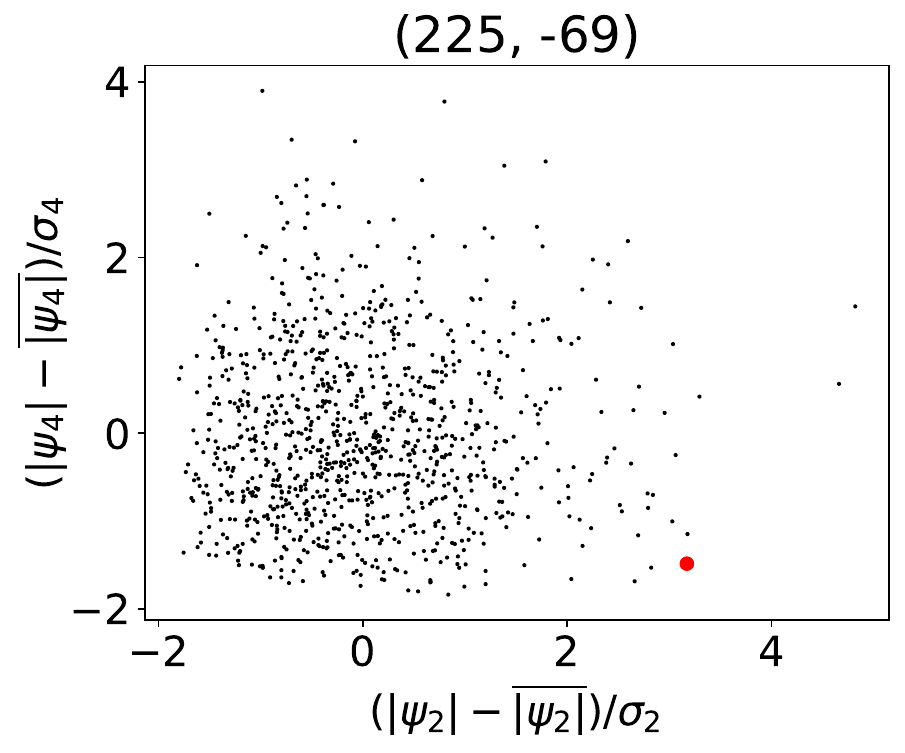} &
    \hspace{-0.2cm}\includegraphics[trim=-0.25cm 0 0.25cm 0, clip, width=0.32\textwidth]{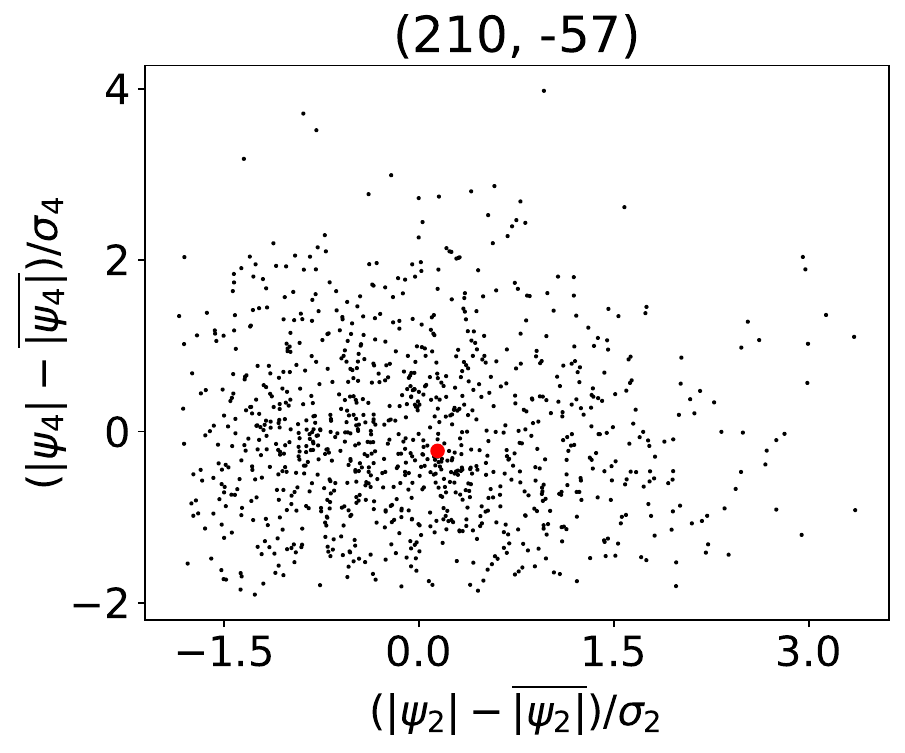}
    \end{tabular}
    \caption{Scatter plots showing irreducible anisotropy at ranks 2 and 4 for
    single MM pixels at (80,\,27) (a), the region next to the Cold Spot
    (225,\,-69) (b), and the Cold Spot at (210,\,-57) (c). Small black dots:
    simulations; large red dots: data}
    \label{fig:scatterplots}
\end{figure}

For a more comprehensive shape analysis of the two anomalous regions and
the Cold Spot, we also performed a multivariate analysis that combines
anisotropy of different ranks. Therefore, we created scatter plots that
show the anisotropy of ranks 2 and 4 for the MM pixels, which are based
on the IMT. To highlight outliers, the scatter plots compare the results
of simulations and data. Figure~\ref{fig:scatterplots} shows one scatter
plot for each region. Importantly, the anisotropy information of the two
different ranks appears to be independent~---~by virtue of our
irreducible representation. For the Cold Spot, the data lies well within
the simulations; for the spot close by, i.e., $(l_1,b_1)=(225,-69)$, the
rank 2 IMT exceed most simulations, but not the rank 4 IMT. Only for
the spot at (80,27) do we find that both rank 2 and 4 IMT are distinctly
larger for the data than for any of the simulations. None of the spots
are noteworthy if we only use $|\psi_4|$. The Cold spot does not appear
noteworthy in the MM because we analyze its shape and not absolute value
information. Hence, our analysis provides information that is
complementary to that of classical methods.

\section*{Discussion}
We propose a new framework for shape analysis on the sphere using
Cartesian and irreducible Minkowski tensors. We, specifically, devised
an irreducible representation on the sphere to gain access to
higher-rank tensors. We, moreover, introduced MM to the sphere,
implementing parallel transport. The MM provide a path for a localized
analysis at any given scale of interest. Adaptation to interdisciplinary
fields of research is facilitated by our open-source software
\texttt{litchi}. Previous open software on the sphere only incorporated
the global MF (e.g., \cite{carones2023,hamann2023}); to our knowledge no
tools for calculating any MT on the sphere are available to the public
and no such local analysis has been done by other groups.
\texttt{litchi} is not only the first available open-source tool for
calculating MF and MT on the sphere using MM, it also provides access to
arbitrary rank via the IMT. 

We applied these tools to CMB data, finding noteworthy spots in CMB
temperature data. One of them is near the Cold Spot, the other above the
Galactic plane. Next, a more elaborated analysis is needed to
disentangle the nature and origin of the anomalies detected here. Such
an analysis is beyond the scope of this paper, where we present the
general framework of Minkowski tensors on the sphere.

Importantly, we provide a generic toolset for anisotropy on the sphere
that can be applied to any spherical data set
with manifold conceivable applications. We mention exemplarily the
characterization of patterns in earth observation data (see, e.g.,
\cite{hersbach20}), for climate research (like teleconnection patterns
induced by El Ni\~{n}o-Southern Oscillation (ENSO) \cite{agarwal19,
strnad22}), heterogeneous pattern on cell surfaces(e.g., patches of
highly adhesive proteins~\cite{spengler_adhesion_2024}), or emergence
phenomena of active matter on the sphere (see, e.g.,
\cite{praetorius18,nitschke19,hsu22,dlamini21}), where the role of a
non-Euclidean geometry on self-organization is not yet understood. 
In these and many other cases, MT as shape descriptors that capture
$n$-point information will likely give new insights into the
phenomenologies and their underlying governing rules.

\section*{Methods}
\subsection*{Basics of Planck data analysis}
Planck data maps are given in the HEALPix (Hierarchical Equal Area
isoLatitude Pixelation) format \cite{healpix}, for which software
packages in various languages are provided at
\url{https://healpix.sourceforge.io/}. A multitude of Planck data
products are available in the Planck Legacy Archive
(\url{https://pla.esac.esa.int}), including component-separated
temperature and polarisation maps generated with different pipelines and
simulations thereof. Detailed descriptions of component separation are
given in \cite{planck2016-l04} and Full Focal Plane (FFP) simulations
are described in \cite{planck2014-a14}. Additional information can be 
found in the Planck Explanatory Supplement (\url{https://wiki.cosmos.esa.int/planck-legacy-archive/index.php/Main_Page}).

The pixel size is characterized by the \Nside\ parameter which gives the
number of pixels as $N_\text{tot} = 12 N_\text{side}^2$. The maximum
resolution available is usually \Nside\ = 2048 (corresponding to a pixel
radius of up to 1.8'). Downgrading to a lower resolution is usually done
by transforming into harmonic space, deconvolving with the previous beam
size, convolving with a larger window, and transforming back to real
space at a lower resolution. Typical convolution FWHMs that are also
used in this paper are given in \cite{planck2016-l07}.

\subsection*{Marching squares on HEALPix}

The marching square procedure gives the tensors in a grid whose pixel
centers are located at the intersections of the original HEALPix grid
pixels and whose corners are at the original pixel centers. The result is not
a HEALPix grid. We implemented this MM-grid by assigning each MM pixel
the number of the HEALPix pixel to its east (so each HEALPix-pixel
number refers to the western corner of that pixel).
This procedure is demonstrated for an exemplary pixel in
Supplementary Fig.~3.
Note that this technique cannot be applied to the poles. Hence, we assigned special
negative numbers to the poles in the MM-grid.

Contours are distorted on the HEALPix grid. This effect is visible when looking
at the average neighbor distance of each pixel
(Supplementary Fig.~4). The distorted pixel shapes affect the contour
length when looking at small, pixel scale structures, but the effect is
less prominent for larger objects. Our tests found that rotating areas of
interest to the origin did not significantly change the deviation
between data and simulations. Since data and simulations are equally
affected, we chose to neglect this effect.  

Parallel transport in \texttt{litchi} is implemented as a single step in
the Schild's ladder procedure (introduced by Alfred Schild in
unpublished lectures at Princeton University, presented in
\cite{gravitation}). A recursive scheme enabling several steps was
tested and made no significant difference over the relevant angular
distances.

Masks can be applied to the image in \texttt{litchi} by giving a
HEALPix-file of the same \Nside\ and numbering scheme as the data to be
analyzed. The mask is expected to take values between zero and one, and
a threshold is applied (default: 0.9). 
Any pixel below this value is assumed to be masked and set to NAN (``not a number'') in the
data. No contours touching a masked area are included in the further
analysis. If more than 1 in 16 pixels contributing to an output MM pixel
are masked, the output pixel is set to NAN.

\subsection*{Rank 4 Anisotropy measures}
On the path to an anisotropy measure on rank 4, we tried several
possibilities.
The first ansatz is to use the norm of a vector containing the
eigenvalues and is shown in Supplementary Fig.~5. The results
look very similar to simply calculating the area of the shape in the
window.

Next, we can look at the three eigenvalues directly. Supplementary Figure
6, Fig.~\ref{fig:rectangles_rank4_mid}, and
Supplementary Fig.~7 show the largest to smallest eigenvalue
of the example shapes, respectively. While the largest eigenvalue again
looks similar to the first ansatz, the other two apparently encode
4-fold symmetry information, reaching higher values for the cross and
square.

To better distinguish between the effects of the smallest and middle
eigenvalue, we generated the same images using a triangle and a
parallelogram. While the smallest eigenvalue reaches high values for these
new shapes compared to the former examples (Supplementary Fig.~8,
bottom), the other one does not react to them in a strong way
(Supplementary Fig.~8, top). 
Note that the diagonal lines are pixelated and not as smooth as shown in
the examples, but this does not introduce a 4-fold symmetry. Based on
these tests, we decide that the middle eigenvalue of $W^{0,4}_1$ as
represented in eq.~\ref{eq:rank4mat} is a useful measure of 4-fold
symmetry.

\section*{Data availability}
The data and simulations used in this work are publicly available in the
Planck Legacy Archive at \url{https://pla.esac.esa.int}. Further data
products were generated with \texttt{litchi} with the parameters stated
in the text.

\section*{Code availability}
The MM generation procedure was implemented in \texttt{litchi}, which is
a lightweight tool written in C++. Python bindings are included to allow
for Python-only analysis scripts and its header-only structure allows
for an easy integration with other C++-based projects. It is available
at \url{https://github.com/ccollischon/litchi}

\printbibliography

\section*{Author contributions}
C.R., M.K., M.S., and C.C. conceived the presented idea.
C.C. designed and implemented the computational framework, analyzed the data, and wrote the first draft.
C.C. and M.K. wrote the manuscript with support from C.R..
C.R. and M.K. supervised the project.
A.B. supervised the Planck data analysis.
All authors reviewed the manuscript.

\section*{Competing interests}
The authors declare no competing interests.

\end{document}


\maketitle

\begin{figure}[h]
    \centering
    \includegraphics[trim=0 1.6cm 0 0, clip, width=\textwidth]{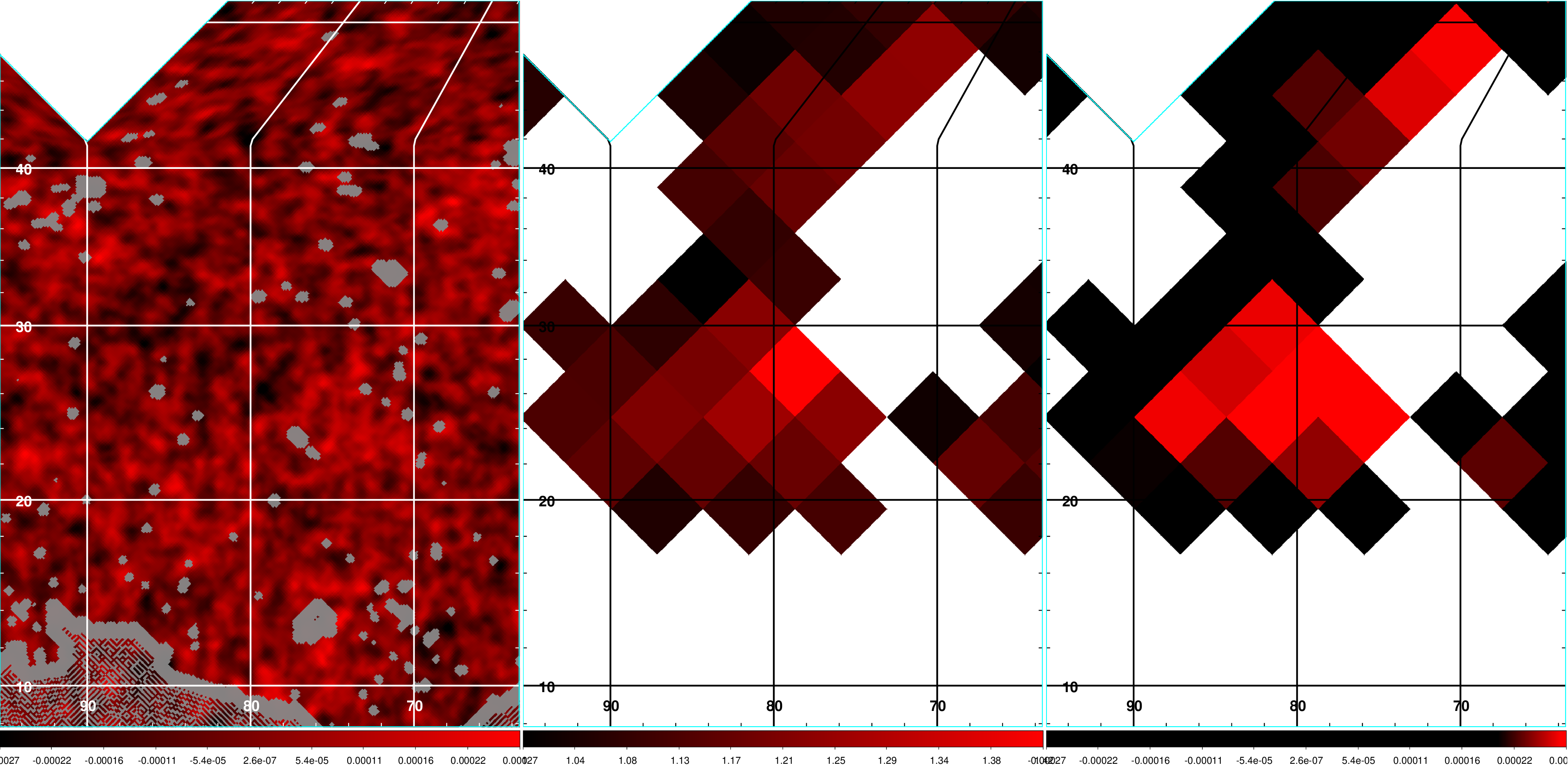}
    \caption{Region surrounding $(l_2, b_2) = (80,27)$. Left: Data at \Nside\ = 512, gray areas are masked. Center: MM of EVQ of $W^{0,2}_1$ for 13 thresholds between -3
and 3 standard deviations with window radius 6\,deg. Right: Local percentile of deviation to simulations. Coordinates are galactic and in deg.}
    \label{fig:8027}
\end{figure}

\begin{figure}
    \centering
    \includegraphics[trim=0 1.6cm 0 0, clip, width=\textwidth]{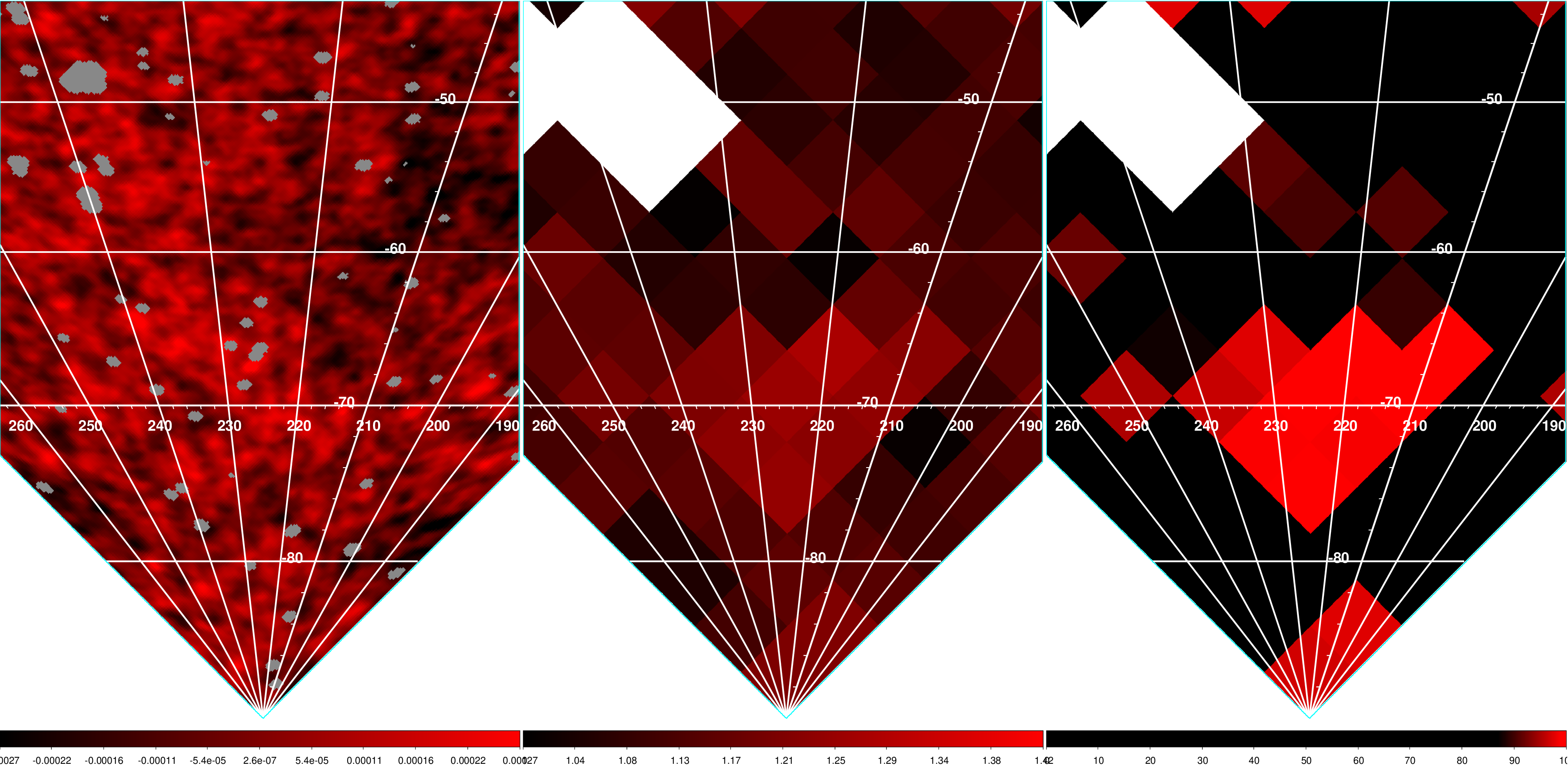}
    \caption{Region surrounding $(l_1, b_1) = (225,-69)$. Left: Data at \Nside\ = 512, gray areas are masked. Center: MM of EVQ of $W^{0,2}_1$ for 13 thresholds between -3
and 3 standard deviations with window radius 6\,deg. Right: Local percentile of deviation to simulations. Coordinates are galactic and in deg.}
    \label{fig:225-69}
\end{figure}

\begin{figure}
\centering
\includegraphics[scale=1]{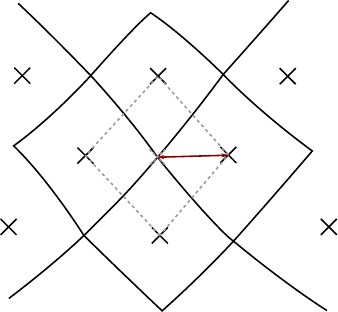}
\caption{Location of a MM pixel within the HEALPix grid. Black lines and
crosses represent pixels in the HEALPix scheme, gray dashed lines and
cross show the boundaries and center of a MM pixel. The number of this
MM pixel would be the same as that of the HEALPix pixel east of it in
its respective map (red line). }
\label{fig:minkmapPixel}
\end{figure}

\begin{figure}
    \centering
    \includegraphics[scale=0.5]{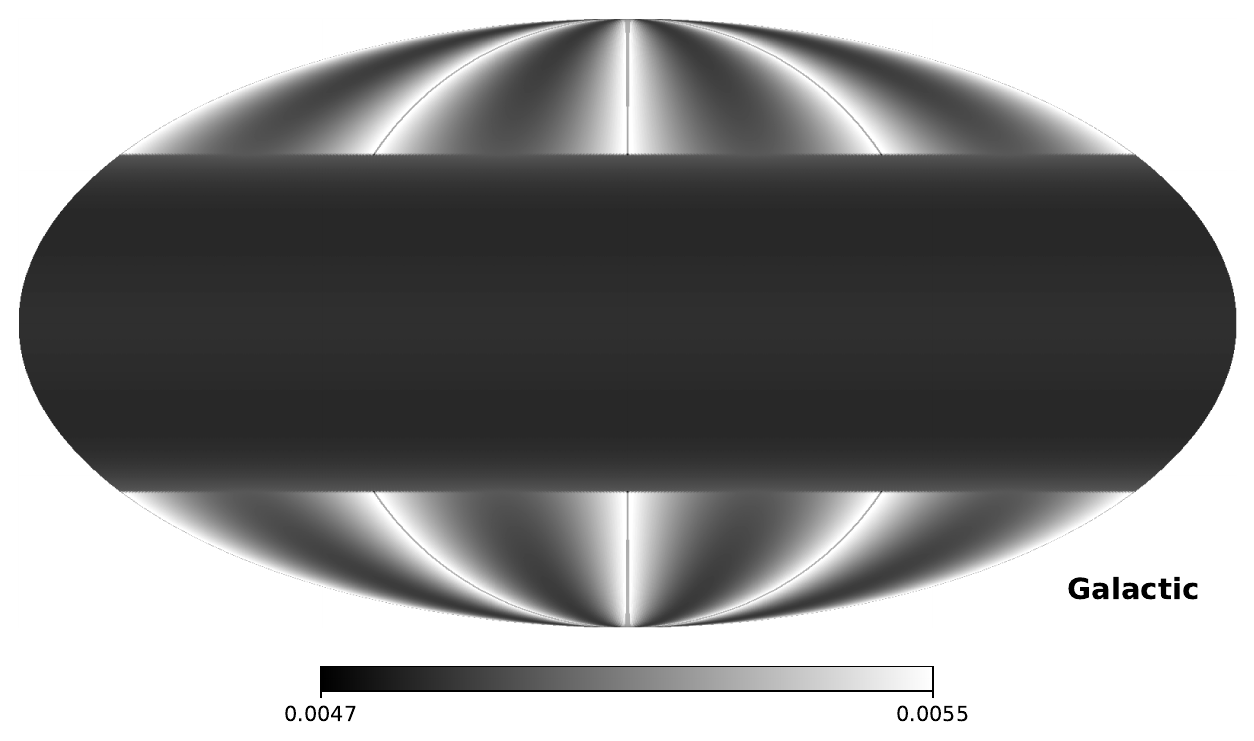}
    \caption{Average distance in radians from each pixel center to its
    neighbors on the HEALPix grid at \Nside\ = 256.}
    \label{fig:averageDist}
\end{figure}

\begin{figure}
    \centering
    \includegraphics[scale=0.9]{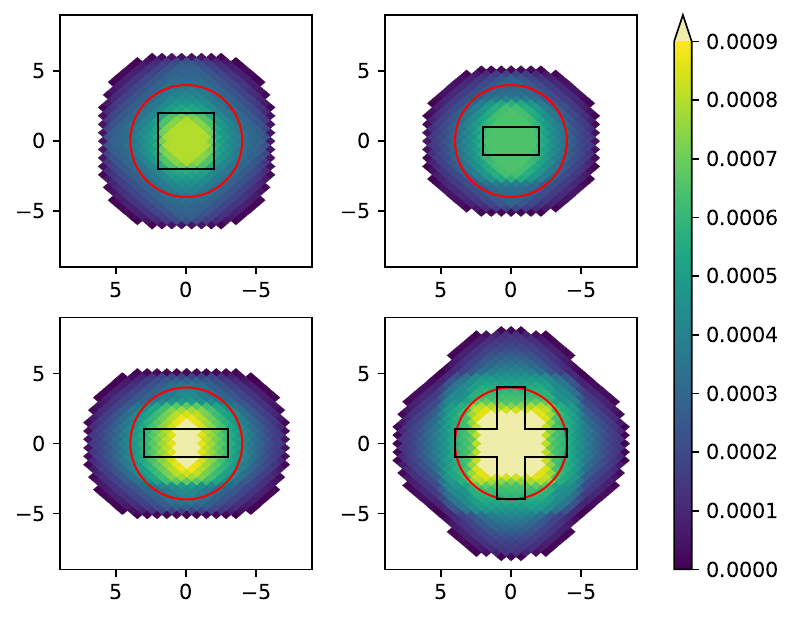}
    \caption{Minkowski maps of the length of the vector of eigenvalues
    of $W^{0,4}_1$ of a square, elongated rectangle, longer rectangle,
    and cross. The black shapes are the shape of the input body, the red
    circles show the window size. Coordinates are given in deg. The value of \Nside\ of
    the input was 512, that of the output is 128.}
    \label{fig:rectangles_rank4_norm}
\end{figure}

\begin{figure}
    \centering
    \includegraphics[scale=0.9]{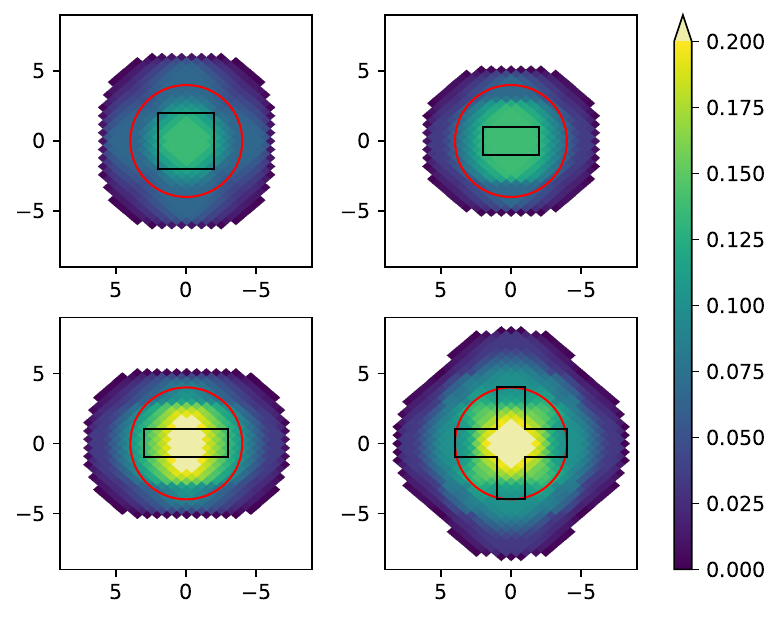}
    \caption{Minkowski maps of the largest eigenvalue of $W^{0,4}_1$ of
    a square, elongated rectangle, longer rectangle, and cross. The
    black shapes are the shape of the input body, the red circles show
    the window size. Coordinates are given in deg. The value of \Nside\ of the input
    was 512, that of the output is 128.}
    \label{fig:rectangles_rank4_max}
\end{figure}

\begin{figure}
    \centering
    \includegraphics[scale=0.9]{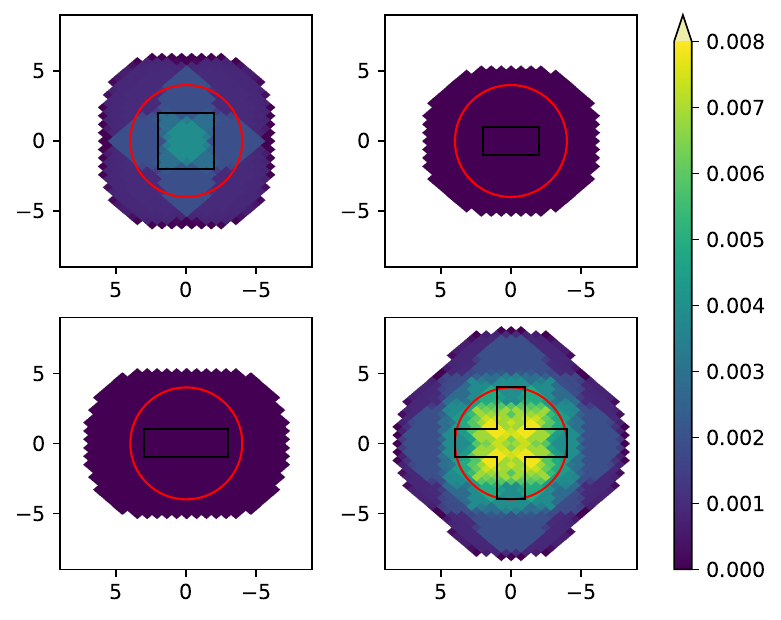}
    \caption{Minkowski maps of the smallest eigenvalue of $W^{0,4}_1$ of
    a square, elongated rectangle, longer rectangle, and cross. The
    black shapes are the shape of the input body, the red circles show
    the window size. Coordinates are given in deg. The value of \Nside\ of the input
    was 512, that of the output is 128.}
    \label{fig:rectangles_rank4_min}
\end{figure}

\begin{figure}
    \centering
    \includegraphics{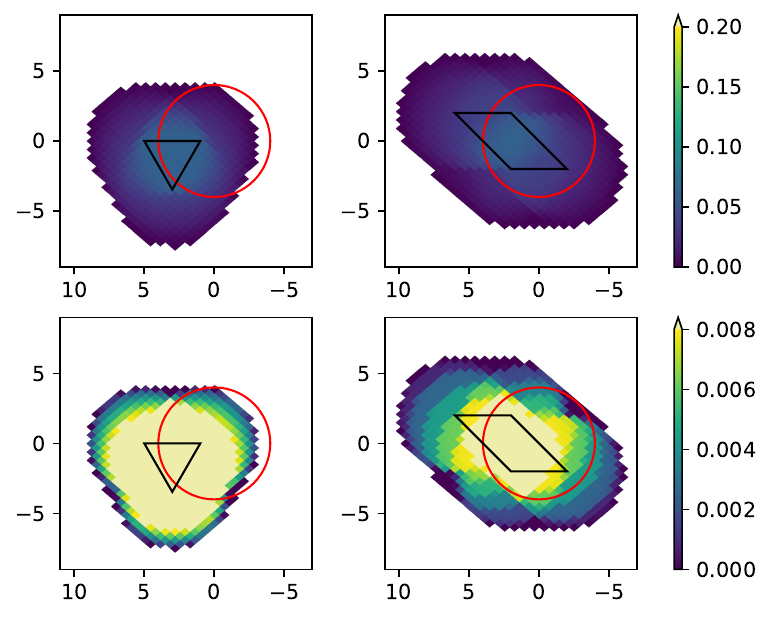}
    \caption{Minkowski maps of the middle (top) and smallest (bottom)
    eigenvalue of $W^{0,4}_1$ of a triangle and parallelogram. The black
    shapes are the shape of the input body, the red circles show the
    window size. Coordinates are given in deg. The value of \Nside\ of the input was
    512, that of the output is 128. Colorbars are the same as in the
    respective images above}
    \label{fig:other_shapes}
\end{figure}